\documentclass[12pt]{article}
\usepackage{epsfig}
\begin{document}
\begin{titlepage}
\begin{center}

\hfill{FTUV-09-1208}

%\hfill{IFIC/09-}

\vspace{2cm}

{\Large \bf Gluonic components of the pion and the transition form factor $\gamma^*\gamma^*\rightarrow\pi^0 $} \\
\vspace{0.50cm}
\renewcommand{\thefootnote}{\fnsymbol{footnote}}
N.I. Kochelev$^{a,b}$\footnote{kochelev@theor.jinr.ru} and V. Vento$^b$
\footnote{vicente.vento@uv.es}
\vspace{0.50cm}

{(a) \it Bogoliubov Laboratory of Theoretical Physics,\\
Joint Institute for Nuclear Research, Dubna, Moscow region, 141980
Russia}
\vskip 1ex
{(b)\it Departament de F\'{\i}sica Te\`orica and Institut de F\'{\i}sica Corpuscular,\\
Universitat de Val\`encia-CSIC, E-46100 Burjassot (Valencia), Spain} \\
\vskip 1ex

\end{center}
\vskip 3.0cm

\centerline{\bf Abstract}
We propose an effective lagrangian for the coupling of the neutral pion with
gluons whose strength is determined by a low energy theorem. We calculate the contribution of the gluonic components arising from this interaction to the pion transition form factor $\gamma^*\gamma^*\rightarrow\pi^0 $ using the instanton liquid model to describe  the QCD vacuum. We find that this contribution is
large and might explain the anomalous  behavior of the form factor at large virtuality
of one of the photons, a feature  which was recently discovered
by the BaBar Collaboration.

\vskip 0.3cm
\leftline{Pacs: 12.38.Lg, 13.40.Gp}
\leftline{Keywords: pion,  gluon component, instanton, form factor}

\vspace{1cm}
\end{titlepage}

\setcounter{footnote}{0}

%\section{Introduction}

The implications, of recent data by the BaBar Collaboration \cite{BaBar} on the transition form factor $\gamma^*\gamma\rightarrow \pi^0 $, in our understanding of the structure of the pion are being
widely discussed  \cite{Dorokhov:2009zx,Radyushkin:2009zg,Polyakov:2009je,Mikhailov:2009sa,chernyak}.
A possible scenario to explain these data consists in assuming a flat
shape for pion distribution amplitude \cite{Radyushkin:2009zg,Polyakov:2009je}, supported by some low energy models like the Nambu-Jona-Lasinio model \cite{RuizArriola:2002bp,Courtoy:2007vy} and instanton based non-local chiral quark model \cite{Anikin:1999cx},and its detailed behavior under Quantum Chromodynamics (QCD) evolution, and a large mass cut-off added to the quark propagator \cite{Radyushkin:2009zg}, signalling a peculiar behavior of the $\overline{q}q$ wave function  \cite{Noguera2009}. Never mind the various explanations, what appears evident is that the BaBar data\cite{BaBar}, if confirmed, are in  contradiction with most  model predictions based on the factorization approach to exclusive reactions at large momentum transfer \cite{Lepage:1979zb} and this apparent lack of perturbative  factorization motivates the present investigation. In this Letter we suggest an alternative nonperturbative explanation to the BaBar results based on the existence of additional contributions to the pion form factor never previously considered. These contributions arise from the admixture of gluonic components, associated to nonperturbative properties of the QCD vacuum, which provide a strong interaction with two photons.

Let us propose a low-energy effective $\pi^0$ interaction with gluons of the following form
\begin{equation}
{\cal{L}}^{eff}_{\pi gg}=-\frac{1}{f_G^{\pi^0}}\pi^0\frac{\alpha_s}{8\pi}G^a_{\mu\nu}\widetilde{G}^a_{\mu\nu}.
\label{lag}
\end{equation}
Such type of Lagrangian density, describing the interaction of a pseudoscalar meson with gluons,
was introduced many years ago by Cornwall and Soni \cite{Cornwall:1984pa} to derive Witten's relation
between the $\eta^\prime$ mass and the topological susceptibility, in a world without light quarks
\cite{Witten:1979vv}
\begin{equation}
\chi_t^{N_f=0}=-\frac{f_\pi^2}{6}(M_{\eta^\prime}^2+M_{\eta}^2-2M_{K}^2),
\label{eta}
\end{equation}
where $f_\pi=92.3$ MeV is the pion decay constant, $M_X$ the mass of the indicated particles, the topological susceptibility is given by
\begin{equation}
\chi_t^{N_f=0}=i\int d^4x <0|T\{Q_5(x)Q_5(0)\}|0>_G,
\label{top}
\end{equation}
and
\begin{equation}
Q_5(x)=\frac{\alpha_s}{8\pi}\, G^a_{\mu\nu}(x)\widetilde{G}^a_{\mu\nu}(x)
\label{topd}
\end{equation}
is the topological charge density.

The effective
pion-gluon interaction, Eq.\ref{lag}, is the analogue of the
pion-quark effective interaction
\begin{eqnarray}
 {\cal
L}^{eff}_{\pi qq}=-\frac{1}{f_\pi}M_q\bar
qi\gamma_5\vec{\tau} q \cdot \vec{\pi}, \label{lagq}
\end{eqnarray}
giving the pion coupling to the quarks.

To derive the decay constant $f_G^{\pi^0}$, which sets the scale of the gluon nonperturbative
interaction with the neutral pion, we will use a low energy theorem (LET)
 \cite{Gross:1979ur}
\begin{equation}
<0|\frac{\alpha_s}{8\pi}G_{\mu\nu}^a\widetilde G_{\mu\nu}^a|\pi^0>
  =\frac{1}{2}\frac{m_d-m_u}{m_d+m_u}f_{\pi}M^2_{\pi}.
  \label{anomaly}
  \end{equation}
 We stress that this matrix element is rather big due to the large
 light quark mass ratio \cite{PDG}
 \begin{equation}
 z=\frac{m_u}{m_d}=0.35-0.6.
 \label{ratio}
 \end{equation}
 By using the effective interaction Eq.\ref{lag} and the LET Eq.\ref{anomaly}
 we get
 \begin{equation}
 f_G^{\pi^0}=-\frac{2(1+z)}{(1-z)}\frac{\chi_t^{N_f=0}}{f_{\pi}^2M^2_{\pi}}f_\pi.
 \label{coupling}
 \end{equation}
 As it was to be expected, the strength of the coupling of the neutral pion to gluons is related
 to the violation of isospin symmetry and proportional
 to the difference of the d- and u-quark masses
\begin{equation}
 \frac{1}{f_G^{\pi^0}}\propto m_d-m_u.
 \label{coup1}
 \end{equation}
 For the value of the mass ratio $m_u/m_d$ shown in Eq.\ref{ratio} which is the one
  allowed by the Particle Data Group,  one obtains
 \begin{equation}
 R=\frac{f_G^{\pi^0}}{f_\pi}\simeq 28.1-51.1.
 \label{coup2}
 \end{equation}

\begin{figure}[h]
\centering
\hspace*{-4cm}
%\vspace*{5cm}
\epsfig{file=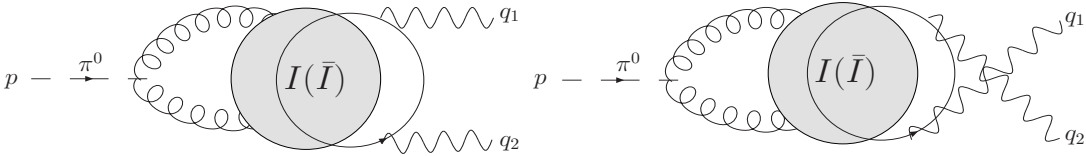,width=14cm,height=18cm,angle=0}
\vspace*{-13cm}
\caption{Gluonic contribution to the  pion transition  form factor. Symbol I ($\bar I$) denotes instanton(antiinstanton). }
\end{figure}

Let us next calculate the contribution arising from the interaction given by Eq.\ref{lag} to the transition form factor of two photons to the pion $\gamma^*(q_1)\gamma^*(q_2)\rightarrow \pi^0 (p)$, where $q_1$ and $q_2$ are the photon momenta and $q_1+q_2=p $. We consider the case where all virtualities of the incoming and the outcoming particles
are in the Euclidean domain $Q_1^2=-q_1^2\geq 0 $, $Q_2^2=-q_2^2\geq 0 $,  $P^2=-p^2\geq 0 $.
The Lagrangian density Eq.\ref{lag} describes the pion interaction with soft gluons. Such gluons should
interact with photons through nonperturbative QCD interactions. We use the instanton liquid model (ILM) for the QCD vacuum to calculate the interaction of two photons with the gluonic component of the pion.
The ILM is one of the most successful models for the description of nonperturbative
QCD effects (see reviews \cite{shuryak,diakonov}).
Within the ILM the single instanton contribution to the pion form factor coming from the interaction Eq.\ref{lag} is associated to the diagrams in Fig.1.

 The  amplitude for the $\pi^0-\gamma^*\gamma^*$ interaction via an instanton with center
  $z_0$ has the following form
\begin{eqnarray}
T_{\mu\nu}(p,q_1,q_2)&=&\epsilon_{\mu\nu\alpha\beta} \sum_i e_i^2\frac{4}{f_G^{\pi^0}\pi^4}\int n_I(\rho)\rho^2d\rho \times \nonumber\\
& & \int d^4z_0\int d^4z_1\int d^4x\int d^4y e^{-ipz_1}e^{iq_1x}e^{iq_2y}Q^I_5(\bar z_1)) \times
 \nonumber\\
&&
\frac{h_{\bar x}h_{\bar y}}{\Delta^2}\left\{\frac{1}{\Delta^2}
(h_{\bar y}\Delta^\alpha\bar y^\beta-h_{\bar x}\Delta^\beta\bar x^\alpha)+
h_{\bar x}h_{\bar y}\bar x^\alpha\bar y^\beta\right\},
\label{amp}
\end{eqnarray}
where
\begin{equation}
Q_5^I(\bar z_1))=\frac{6\rho^4}{\pi^2(\bar z_1^2+\rho^2)^4}
\label{topcharge}
\end{equation}
is the topological charge density of the instanton in the space-time point $z_1$,
 $n_I(\rho)$ is the instanton density, $\rho$ is the instanton size, $\Delta=\bar x-\bar y$,
$h_{\bar x}=1/(\bar x^2+\rho^2)$, $h_{\bar y}=1/(\bar y^2+\rho^2)$ and the notation $\bar w\equiv w-z_0$ for any variable $w$ has been introduced. The sum runs over the light quark flavors, i.e.  $i=u,d,s$.
To get Eq.\ref{amp} we have used  the correlator of two electromagnetic
currents in the instanton field obtained by Andrei and Gross \cite{Andrei:1978xg}. Their result was corrected by a color factor (see \cite{shuryak2}). The first term in the last line of the equation is coming from the quark nonzero modes in the instanton field and the last term arises from the interference between nonzero and zero modes \cite{Andrei:1978xg}.

The final result for the gluon contribution to the pion transition form factor induced by instantons
is
\begin{equation}
F(P^2,Q_1^2,Q_2^2)^I_g=\frac{4<e^2>}{f_\pi R}
\int d\rho n_I(\rho)\rho^4S(\rho,P^2,Q_1^2,Q_2^2)
\label{final}
\end{equation}
where
\begin{equation}
S(\rho,P^2,Q_1^2,Q_2^2)=\Phi_1(\sqrt{z_3})\int_0^1dt\left\{I(t,z_1,z_2,z_3)+(1-t)I(t,z_2,z_1,z_3)\right\},
\label{ss}
\end{equation}
\begin{eqnarray}
I(t,z_1,z_2,z_3)&=& \int_0^{\infty}d\alpha\frac{\alpha(\alpha+1)\Phi_2(Z(\alpha,t,z_1,z_2,z_3))}{(\alpha+1-t)^3
Z^2(\alpha,t,z_1,z_2,z_3)},
\label{int}
\end{eqnarray}
and
\begin{equation}
Z(\alpha,t,z_1,z_2,z_3)=\sqrt{(\alpha+1)(t\alpha z_1+tz_2+(1-t)z_3)/(\alpha+1-t)}.
\label{z}
\end{equation}
The functions
\begin{equation}
\Phi_1(z)=\frac{z^2K_2(z)}{2}, \; \Phi_2(z)=zK_1(z)
\label{for2}
\end{equation}
behave as $\Phi_{1,2}(z)\rightarrow 1$ in the limit $z\rightarrow 0$. In Eqs.\ref{final}-\ref{z} the notations are
  $z_1=Q_1^2\rho^2$, $z_2=Q_2^2\rho^2$, $z_3=P^2\rho^2$ and $<e^2>=\sum_i e_i^2$.

For an estimate we use Shuryak`s version of the ILM \cite{shuryak3},
where the density is given by
\begin{equation}
n_I(\rho)=n_0\delta(\rho-\rho_c)
\label{density}
\end{equation}
and
\begin{equation}
n_0\approx 1/2fm^{-4}, \ \ \  \rho_c\approx 1/3fm.
\label{par}
\end{equation}
Within this simple model for the instanton distribution the result for the form factor is
\begin{equation}
F(P^2,Q_1^2,Q_2^2)^I_g=\frac{4<e^2>f_I}{\pi^2f_\pi R}
S(\rho_c,P^2,Q_1^2,Q_2^2),
\label{final2}
\end{equation}
where $f_I=\pi^2n_0\rho_c^4$ is so-called instanton packing fraction in the QCD vacuum.

It should be  pointed out  that in spite of the smallness of instanton packing fraction
 $f_I\approx 0.06$,  using the single instanton approximation as above is only  valid
for values of the momentum transfers  $Q_1,Q_2\gg 1/R_I$, where $R_I\approx 3\rho_c$ is the distance
between the instantons in the ILM. For smaller photon virtualities it is necessary  to include
the contributions arising from multiinstanton configurations.
With an average size of the instanton in the QCD vacuum as in Eq.\ref{par}
for the region  $Q_1^2,Q_2^2\geq 1/\rho_c^2 \geq \mu^2=0.35$ GeV$^2$, i.e. $z_{1,2}\geq 1 $,
the validity of a single instanton approximation is assured.

The calculation above was done for the case when all external momenta are Euclidean. In order to compare with BaBar data we have to perform an analytic continuation of the pion virtuality to the physical point of the pion on-shell $P^2\rightarrow-m_\pi^2-i\epsilon$.
An inspection of the integrals in Eqs.\ref{ss},\ref{int} shows that the dominant contribution to the form factor
at $m_\pi^2/Q_{1,2}^2\ll 1$ is coming from the region of integration $t\approx 0$, due to the pole at $Z^2=0$.
Assuming the  following behavior of the function $\Phi_2(Z)\equiv ZK_1(Z)\sim 1$  near the pole and keeping
only leading terms in $m_\pi^2$  we obtain  following closed form
formulas for the real and imaginary parts of the flavor singlet part of form factor

\begin{eqnarray}
Re(F(m_\pi^2,Q_1^2,Q_2^2)^I_g)&\simeq&\frac{4<e^2>f_I}{\pi^2f_\pi R}\times\nonumber\\
&& \{ \frac{z_1[z_2log(z_2)/z_1+log(z_1)(log(z_1/z_2)-1)+Li_2((z_1-z_2)/z_1)]}{(z_1-z_2)^2}+ \nonumber\\
&&\frac{z_2[z_1log(z_1)-\pi^2/6-log^2(z_1-z_2)/2 - log(z_2)(1-log(z_1)/2)]}{(z_1-z_2)^2}+ \nonumber\\
&&\frac{z_2[Log(z_2)log((z_1/(z_1-z_2))-Li_2(z_2/(z_2-z_1))]}{(z_1-z_2)^2}-\nonumber\\
&& \frac{log(z_1/z_2)log(m_\pi^2\rho_c^2)}{z_1-z_2}
\},
\label{real}\\
 Im(F(m_\pi^2,Q_1^2,Q_2^2)^I_g)&\simeq&\frac{4<e^2>f_I}{\pi f_\pi R}\frac{ log(z_1/z_2)}{z_1-z_2}.
\label{im}
\end{eqnarray}
The imaginary part of form factor arises because the pion may decay in this calculation into a quark-antiquark pair since confinement, which forbids this decay, is not explicitly implemented. However, the net contribution of the imaginary part to total transition form factor in the BaBar kinematics is very small. These formulas are useful to extract the behavior of the transition form factor with $Q^2$. The exact numerical analysis will be described below. For definiteness we consider the case $z_1>z_2$

 In the limit $Q_1^2\gg Q_2^2$ which is valid for BaBar kinematics, the formulas for the
 real and the imaginary parts  simplify,
 \begin{eqnarray}
 Re(F(m_\pi^2,Q_1^2,Q_2^2))^I_g &\approx& \frac{4<e^2>f_I}{\pi^2f_\pi R}
 \frac{[log(Q_1^2/m_\pi^2)log(Q_1^2/Q_2^2)+\pi^2/6]}{\rho_c^2Q_1^2},\label{re2}\\
 Im(F(m_\pi^2,Q_1^2,Q_2^2))^I_g &\approx& \frac{4<e^2>f_I}{\pi f_\pi R}\frac{log{(Q_1^2/Q_2^2)}}{\rho_c^2Q_1^2}.
 \label{im2}
 \end{eqnarray}
 It follows from Eqs.\ref{re2},\ref{im2} that the flavor singlet gluon induced part of the form factor has a dependence on the large photon virtuality $Q_1^2$ proportional to $log^2(Q_1^2)/Q_1^2$, which is much stronger than that of the flavor nonsinglet part, which in most of the models is of the form
 $1/Q_1^2$. The additional feature of this new contribution is its  strong chiral enhancement since the massless logs appear governed by the pion mass as $log(Q_1^2/m_\pi^2)$.
 For symmetric kinematics $Q^2=Q_1^2=Q_2^2$ the result is

\begin{eqnarray}
 Re(F^S(Q^2))^I_g &\approx& \frac{<e^2>f_I}{\pi^2f_\pi R}
 \frac{(3+2log(Q^2/m_\pi^2))}{\rho_c^2Q^2},\label{re3}\\
 Im(F^S(Q^2))^I_g &\approx& \frac{2<e^2>f_I}{\pi f_\pi R\rho_c^2Q^2}.
 \label{im3}
 \end{eqnarray}

Having determined the dependence of the virtuality in the approximation Eq. \ref{real}, we proceed to study the exact numerical calculation which includes the effect of the functional form of
$ZK_1(Z)$. Before we do so, we would like to point out that the exact calculation leads to a smaller (by about 50\%) result, compared to the approximate calculation. This factor can be absorbed in the  uncertainties of the vacuum model associated with the poor knowledge of the instanton distribution (about 30\%), and the additional uncertainty coming from the  value of the pion coupling to gluons Eqs.\ref{coup1} and \ref{coup2} (about a factor 2) due indeterminacy in the ratio of  u- and d- quark masses, Eq.\ref{ratio}.

We compare our result  with the BaBar data. Before we do so some caveats have to be expressed since the comparison is not direct. The BaBar experiment, only measures the virtuality of one of the photons in the interval $Q_1^2=4-40$ GeV$^2$. They only put an upper limit on the virtuality for the second photon,
$Q_2^2<0.18 $ GeV$^2$. Finally, they use a  model for the form factor
 to extract the value at the real photon point $Q_2^2=0$. Thus, a direct comparison of our results with the BaBar data is not possible. Moreover, our calculation only represents  the flavor singlet contribution to the form factor, therefore we have to add a flavor nonsinglet part. We take in the estimate shown in Fig.2 for the flavor nonsinglet part the corresponding to a vector meson dominance (VMD) model, i.e.
\begin{equation}
F(Q_1^2,Q_2^2)_q^{VMD}=\frac{1}{4\pi^2f_\pi}\frac{1}{(1+Q_1^2/M_\rho^2)(1+Q_2^2/M_\rho^2)}.
\label{VMD}
\end{equation}

In order to compare our results with the  BaBar data we perform an extrapolation of their results
from   $Q_2^2=0$ to  $Q_2^2=0.35$ GeV$^2$. In Fig.2 we compare  our calculation with  the extrapolation of the BaBar data described by
\cite{BaBar}
\begin{equation}
Q^2\mid F^{BaBar}_{exp}(Q^2)\mid=A\biggl(\frac{Q^2}{10 GeV^2}\biggr)^{\beta},
\label{exp}
\end{equation}
where $A\simeq0.182$ and $\beta\simeq 0.25$. This function has been continued to the point $\mu^2$(0.35 GeV$^2$ in our case) following the VMD model,
\begin{equation}
Q^2\mid F^{BaBar}(Q^2,\mu^2)\mid=\frac{Q^2\mid F^{BaBar}_{exp}(Q^2)\mid}{1+\mu^2/M_\rho^2}.
\label{babar1}
\end{equation}
The bands in the figure represent our uncertainties, both in the vacuum model and in the coupling constant, as mentioned before.

 \begin{figure}[h]
\centering
%\hspace*{-6cm}
%\vspace*{5cm}
\epsfig{file=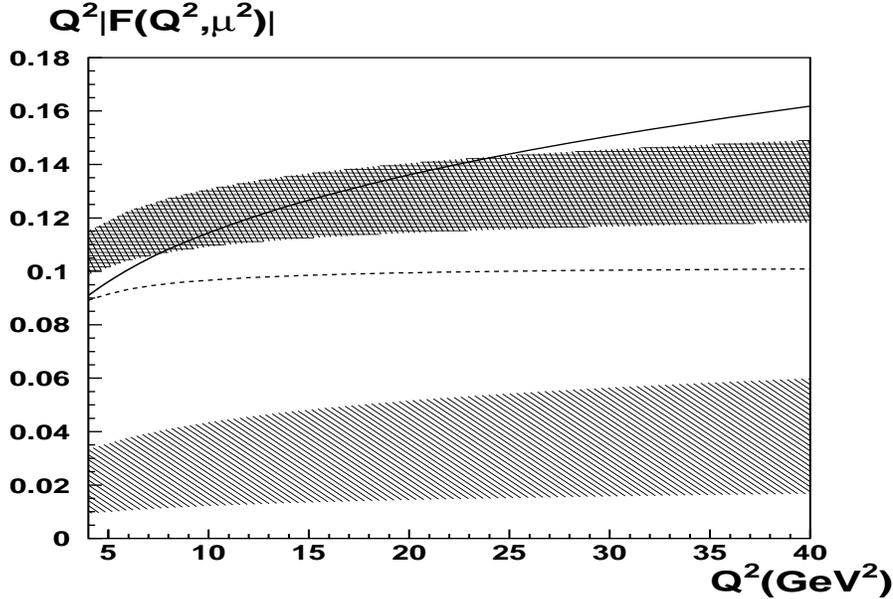,width=12cm,height=8cm,angle=0}
%\vspace*{-10cm}
\caption{ Contributions to the pion transition form factor compared with the  extrapolation of the BaBar data to $Q_2^2=\mu^2=0.35$ GeV$^2$ (solid line): gluonic contribution in our model including uncertainties (lower band), conventional VMD contribution (dashed line) line, and their sum (upper band). }
\end{figure}
It should be mentioned
that the behavior of the gluonic part of the form factor as function of $Q^2$
is determined by  shape of the decay of the non-zero modes in the instanton field (see Eq.\ref{amp}). At the same time it is well known that the VMD-like behavior of the flavor nonsinglet part of pion form factor
can be easily reproduced within a non local chiral quark model based on the quark zero-modes dominance in the instanton vacuum\cite{Dorokhov:2002iu}. Due to  the weaker decay of the quark nonzero modes with respect to  zero modes one can expect  a harder $Q^2$ dependence of the flavor singlet part of the form factor in comparison with the flavor nonsinglet part. Such
tendency is seen in Fig.2. Indeed,
the gluonic part is described well by a fast increasing function of $Q^2$, Eq.\ref{re2}. That function
 shows a similar behavior, as function of $Q^2$, as the BaBar data, Eq.\ref{exp}. Contrary to the gluonic part of the form factor, its flavor nonsinglet
part has a conventional $1/Q^2$ behavior at large $Q^2$, Fig.2.
Taking into account some uncertainties in our estimates related to the poorly known ratio of the u- and d- quark masses,
Eq.\ref{ratio}, as well as uncertainties in the parameters of the instanton model,
 we conclude, that the new contribution related to the gluonic component of pion  might explain the  anomalous behavior of the pion transition form factor found by the BaBar Collaboration. One should be aware that these contributions are beyond the Operator Product Expansion and their QCD evolution is non trivial \cite{Manohar:1990hu}.

 It is evident that such type of contribution
 should be present also for the $\eta$ and $\eta^\prime$ mesons. In this case one should carefully take into account the effects
 of their larger masses  and their strong mixing
 (see recent discussion in the papers \cite{Mathieu:2009sg, Klopot:2009cm}). We also would like to point out that the
 strong $Q^2$ dependence of the gluonic part of the form factor  opens a new possibility  to disentangle
 particles with dominant gluonic  content, i.e. glueballs (see  review \cite{Mathieu:2008me}), in $\gamma^*\gamma^*$ collision.
 These tasks are the subject of future investigations.

In summary, the BaBar data \cite{BaBar} point towards a breaking of perturbative factorization. This has led us to investigate possible non perturbative mechanism that contribute to the pion transition form factor. We have shown that a nonzero interaction of neutral pion with gluons
arising from isospin violation, i.e.  $m_u\neq m_d$, induces a large contribution to this
form factor at large virtuality of one of the photons. More
sophisticated models  for the instanton density and probably multiinstanton
contribution might bring the value closer to the observed one.

\section*{Acknowledgments}
We are grateful to A.E. Dorokhov and S. Noguera,  for very useful discussions. NIK thanks the Departament de F\'{\i}sica Te\`orica Universitat de Val\`encia  for the hospitality and the University of Valencia for a Visiting Professor appointment. This work was supported in part  by HadronPhysics2,  a FP7-Integrating Activities and Infrastructure Program of the European Commission under Grant 227431, by the MICINN (Spain) grant FPA2007-65748-C02-01, by GVPrometeo2009/129, by the RFBR grant 10-02-00368-a and by  Belarus-JINR grant.


\begin{thebibliography}{99}

\bibitem{BaBar}
  B.~Aubert {\it et al.}  [The BABAR Collaboration],
  %``Measurement of the gamma gamma* --> pi0 transition form factor,''
  Phys.\ Rev.\  D {\bf 80} (2009)  052002.

  \bibitem{Dorokhov:2009zx}
  A.~E.~Dorokhov,
  %``Rare decay \pi^0 \to e^+e^- as a Test of Standard Model,''
  arXiv:0905.4577 [hep-ph];
  %``Muon g-2, rare decays P \to l^+l^- and transition form factors P \to
  %\gamma\gamma^*,''
  arXiv:0909.5111 [hep-ph].

\bibitem{Radyushkin:2009zg}
  A.~V.~Radyushkin,
  %``Shape of Pion Distribution Amplitude,''
  arXiv:0906.0323 [hep-ph].

  \bibitem{Polyakov:2009je}
  M.~V.~Polyakov,
  %``On the Pion Distribution Amplitude Shape,''
  JETP Lett.\  {\bf 90} (2009) 228
  [arXiv:0906.0538 [hep-ph]].

  \bibitem{Mikhailov:2009sa}
  S.~V.~Mikhailov and N.~G.~Stefanis,
  %``Transition form factors of the pion in light-cone QCD sum rules with
  %next-to-next-to-leading order contributions,''
  Nucl.\ Phys.\  B {\bf 821} (2009) 291
  [arXiv:0905.4004 [hep-ph]];
  %``Pion transition form factor at the two-loop level vis-\`a-vis experimental
  %data,''
  arXiv:0910.3498 [hep-ph].

\bibitem{chernyak}
  V.~L.~Chernyak,
  %``Exclusive \gamma*\gamma processes,''
  arXiv:0912.0623 [hep-ph].


%\cite{RuizArriola:2002bp}
\bibitem{RuizArriola:2002bp}
  E.~Ruiz Arriola and W.~Broniowski,
  %``Pion light-cone wave function and pion distribution amplitude in the
  %Nambu-Jona-Lasinio model,''
  Phys.\ Rev.\  D {\bf 66} (2002) 094016
  [arXiv:hep-ph/0207266].
  %%CITATION = PHRVA,D66,094016;%%

%\cite{Courtoy:2007vy}
\bibitem{Courtoy:2007vy}
  A.~Courtoy and S.~Noguera,
  %``The Pion-Photon Transition Distribution Amplitudes in the Nambu-Jona
  %Lasinio Model,''
  Phys.\ Rev.\  D {\bf 76} (2007) 094026
  [arXiv:0707.3366 [hep-ph]].
  %%CITATION = PHRVA,D76,094026;%%

\bibitem{Anikin:1999cx}
  I.~V.~Anikin, A.~E.~Dorokhov and L.~Tomio,
  %``On high Q**2 behavior of the pion form-factor for transitions gamma*  gamma

  %--> pi0 and gamma* gamma* ---> pi0 within the nonperturbative  approach,''
  Phys.\ Lett.\  B {\bf 475} (2000) 361
  [arXiv:hep-ph/9909368].

  \bibitem{Noguera2009} S. Noguera and V. Vento, work in preparation.



\bibitem{Lepage:1979zb}
  G.~P.~Lepage and S.~J.~Brodsky,
  %``Exclusive Processes In Quantum Chromodynamics: Evolution Equations For
  %Hadronic Wave Functions And The Form-Factors Of Mesons,''
   Phys.\ Lett.\  B {\bf 87} (1979) 359;
  %``Exclusive Processes In Perturbative Quantum Chromodynamics,''
  Phys.\ Rev.\  D {\bf 22} (1980) 2157.






\bibitem{Cornwall:1984pa}
  J.~M.~Cornwall and A.~Soni,
  %``Couplings Of Low Lying Glueballs To Photons And To Heavy Quarks,''
  Phys.\ Rev.\  D {\bf 32} (1985) 764.

\bibitem{Witten:1979vv}
  E.~Witten,
  %``Current Algebra Theorems For The U(1) Goldstone Boson,''
  Nucl.\ Phys.\  B {\bf 156} (1979) 269.

\bibitem{Gross:1979ur}
  D.~J.~Gross, S.~B.~Treiman and F.~Wilczek,
  %``Light Quark Masses And Isospin Violation,''
  Phys.\ Rev.\  D {\bf 19} (1979) 2188.


\bibitem{PDG} C. Amsler et al. (Particle Data Group), Phys. Lett. B {\bf 667} (2008) 1,
and 2009 partial update for the 2010 edition.


\bibitem{shuryak} T. Sch\"afer and E.V. Shuryak,
Rev. Mod. Phys. {\bf 70} (1998) 1323.

\bibitem{diakonov} D. Diakonov, Prog. Par. Nucl. Phys.
{\bf 51} (2003) 173.



\bibitem{Andrei:1978xg}
  N.~Andrei and D.~J.~Gross,
  %``The Effect Of Instantons On The Short Distance Structure Of Hadronic
  %Currents,''
  Phys.\ Rev.\  D {\bf 18} (1978) 468.


\bibitem{shuryak2}
  E.~V.~Shuryak,
  %``Theory And Phenomenology Of The QCD Vacuum. 5. Correlators And Sum Rules.
  %The Applications,''
  Phys.\ Rept.\  {\bf 115} (1984) 151.



 \bibitem{shuryak3}
  E.~V.~Shuryak,
  %``The Role Of Instantons In Quantum Chromodynamics. 1. Physical Vacuum,''
  Nucl.\ Phys.\  B {\bf 203} (1982) 93;116;140.



\bibitem{Dorokhov:2002iu}
  A.~E.~Dorokhov,
  %``Pion distribution amplitudes within the instanton model of QCD vacuum,''
  JETP Lett.\  {\bf 77} (2003) 63
  [Pisma Zh.\ Eksp.\ Teor.\ Fiz.\  {\bf 77} (2003) 68]


 \bibitem{Manohar:1990hu}
  A.~V.~Manohar,
  %``THE DECAYS Z $\to$ W pi AND Z $\to$ gamma pi,''
Phys.\ Lett.\  B {\bf 244} (1990) 101.



\bibitem{Mathieu:2009sg}
  V.~Mathieu and V.~Vento,
  %``Pseudoscalar glueball and $\eta-\eta^\prime$ mixing,''
  arXiv:0910.0212 [hep-ph].

\bibitem{Klopot:2009cm}
  Y.~N.~Klopot, A.~G.~Oganesian and O.~V.~Teryaev,
  %``Dispersive Approach to Abelian Axial Anomaly, Mixing of Pseudoscalar Mesons
  %and Symmetries,''
  arXiv:0911.0180 [hep-ph].


\bibitem{Mathieu:2008me}
  V.~Mathieu, N.~Kochelev and V.~Vento,
  %``The Physics of Glueballs,''
  Int.\ J.\ Mod.\ Phys.\  E {\bf 18} (2009) 1 [arXiv:0810.4453 [hep-ph]].



\end{thebibliography}
\end{document}